\newcounter{myctr}
\def\myitem{\refstepcounter{myctr}\bibfont\noindent\ifnum\themyctr>9\else\phantom{0}\fi\hangindent17pt\themyctr.\enskip}

\documentclass{ws-ijqi}

\newcommand{\ket}[1]{{|#1\rangle}}
\newcommand{\bra}[1]{{\langle#1|}}
\newcommand{\braket}[2]{{\langle#1|#2\rangle}}
\newcommand{\ketbra}[2]{| #1 \rangle\langle #2 |}

\renewcommand{\i}{\textrm{i}}

\newcommand{\abs}[1]{| #1 |}

\renewcommand{\Re}{\mathrm{Re}}
\renewcommand{\Im}{\mathrm{Im}}

\newcommand{\zon}{\{0,1\}^n}

\newcommand{\be}{\begin{equation}}
\newcommand{\ee}{\end{equation}}
\newcommand{\bea}{\begin{eqnarray}}
\newcommand{\eea}{\end{eqnarray}}

\newcommand{\id}{I}
\def\opone{\leavevmode\hbox{\small1\kern-3.8pt\normalsize1}}

\begin{document}

\markboth{LAWRENCE M. IOANNOU and MICHELE MOSCA} {LIMITATIONS OF
SOME SIMPLE ADIABATIC QUANTUM ALGORITHMS}

\catchline{}{}{}{}{}

\title{LIMITATIONS OF SOME SIMPLE ADIABATIC QUANTUM ALGORITHMS}

\author{LAWRENCE M. IOANNOU}

\address{Department of Applied Mathematics and
Theoretical Physics, University of Cambridge, Wilberforce Road\\
Cambridge, Cambridgeshire, CB3 0WA, United Kingdom\\
lmi22@cam.ac.uk}

\author{MICHELE MOSCA}

\address{Institute for Quantum Computing,
University of Waterloo, 200 University Avenue West\\
Waterloo, Ontario,
N2L 3G1, Canada\\
mmosca@iqc.ca}

\address{Perimeter Institute for Theoretical Physics,
31 Caroline Street North\\
Waterloo, Ontario, N2L 2Y5, Canada}

\maketitle

\begin{history}
\received{Day Month Year}
\revised{Day Month Year}
\end{history}

\begin{abstract}
Let $H(t)=(1-t/T)H_0 + (t/T)H_1$, $t\in [0,T]$, be the Hamiltonian
governing an adiabatic quantum algorithm, where $H_0$ is diagonal in
the Hadamard basis and $H_1$ is diagonal in the computational basis.
We prove that $H_0$ and $H_1$ must each have at least two large
mutually-orthogonal eigenspaces if the algorithm's running time is
to be subexponential in the number of qubits.  We also reproduce the
optimality proof of Farhi and Gutmann's search algorithm in the
context of this adiabatic scheme; because we only consider initial
Hamiltonians that are diagonal in the Hadamard basis, our result is
slightly stronger than the original.
\end{abstract}


\section{Introduction}

\noindent Farhi \emph{et al.}\cite{FGGS00} proposed a
continuous-time quantum algorithm for solving NP-complete
combinatorial search problems, invoking the adiabatic
approximation.\cite{M76} It has been shown that Farhi \emph{et
al.}'s original proposal for the 3SAT problem has exponential
worst-case complexity.\cite{VV02} However, the average-case
performance of (possible modifications of) the original proposal is
still an open question.\cite{FGG00,FGG+01,FGG02,H03}  A more general
notion of adiabatic quantum computation (not considered here) is
universal.\cite{AhaEtAl}

This work\footnote{previously appearing in Ref.~\refcite{Ioa02}}~
supplements the similar recent work\cite{FGGN05,WY06,ZH06} on
analytic lower bounds on the runtime of certain classes of simple
adiabatic quantum algorithms.  Our assumptions are weaker: for our
general bound, we do not assume that the initial Hamiltonian is a
one-dimensional projector.  Thus, the dimension of the
second-largest eigenspace of the Hamiltonians emerges as a factor in
the algorithms' complexity.

Let the function $f: \lbrace 0,1 \rbrace ^{n} \longrightarrow
\lbrace 0,1 \rbrace$ such that
\begin{eqnarray}
f(z)=    \begin{cases}
        1& \textrm{if $z \neq w$}\\
        0& \textrm{if $z=w$}
    \end{cases}
\end{eqnarray}
encode a general search problem (with unique solution $w$) for which
we seek a quantum algorithm. The uniform-amplitude superposition
state,
\begin{eqnarray}\ket{u}
\equiv \frac{1}{\sqrt{N}}\sum_{z\in\{0,1\}^n}\ket{z},
\end{eqnarray}
will be the starting state (at $t=0$) for all the algorithms we
consider. Define the \emph{Walsh-Hadamard transform} (or just
\emph{Hadamard transform})
\begin{eqnarray}
W = \frac{1}{\sqrt{N}} \sum_{x\in\{0,1\}^n} \sum_{y\in\{0,1\}^n}
(-1)^{x\bullet y}\ketbra{x}{y}.
\end{eqnarray}
Let
\begin{eqnarray}
\nonumber \ket{\overline{z}}\equiv W\ket{z},\hspace{10mm}\textrm{for
all $z\in\zon$.}
\end{eqnarray}
We define the \emph{Hadamard basis} to be
$\{\ket{\overline{z}}:z\in\zon\}$.

Recall from Ref.~\refcite{VMV01} and Ref.~\refcite{RC02} that the
continuous-time version of the general search problem, defined in
Ref.~\refcite{FG96}, may be solved adiabatically with initial and
final Hamiltonians
\begin{equation}
H_u = \sum_{z\in \{0,1\}^n}
h_{\textrm{search}}(z)\ketbra{\bar{z}}{\bar{z}}
\end{equation}
and
\begin{equation}
H_w = \sum_{z\in \{0,1\}^n} f(z)\ketbra{z}{z},
\end{equation}
respectively, where $h_{\textrm{search}}(z)\in\{0,1\}$ and
$h_{\textrm{search}}(z)=0 \Leftrightarrow z=0^n$.

If $f$ is the truth function corresponding to a Boolean formula
(also denoted $f$) which is an instance of the 3SAT
problem\cite{GJ79}, then we can define the function $v:\lbrace
0,1\rbrace^n \longrightarrow \lbrace 0,1,\ldots\rbrace$,
\begin{equation}
v(z) \equiv \textrm{the number of clauses of $f$ violated by
assignment $z$}.
\end{equation}
Recall that Farhi \emph{et al.}'s original adiabatic algorithm for
3SAT uses initial and final Hamiltonians
\begin{equation}
H_\textrm{I}=\sum_{z\in\lbrace
0,1\rbrace^n}h(z)\ketbra{\bar{z}}{\bar{z}},
\end{equation}
and
\begin{equation}
H_F=\sum_{z\in\lbrace 0,1\rbrace^n}v(z)\ketbra{z}{z},
\end{equation}
respectively, where $h(z)=\sum_{i:z_i=1}d_i$ and $d_i$ is the number
of clauses of $f$ that contain variable $z_i$.

\section{Results}

\noindent It is convenient to cast Farhi \emph{et al.}'s adiabatic
3SAT algorithm in a more general notation.  Let $N=2^n$.  Let $p(n)$
and $q(n)$ be two nonnegative integer functions of $n$. Let
$\mathcal{P}=\{{P}_j\}_{j=0,1,\ldots,p(n)}$ and
$\mathcal{Q}=\{{Q}_k\}_{k=0,1,\ldots,q(n)}$ be two partitions of
$\{0,1\}^n$, that is, the following two unions are disjoint:
\begin{equation}
\bigcup_{k=0}^{q(n)}Q_{k}=\{0,1\}^n=\bigcup_{j=0}^{p(n)}P_{j}.
\end{equation}
Define the projectors
\begin{eqnarray}
\hat{Q}_{k} \equiv \sum_{z \in Q_k
}\ketbra{\overline{z}}{\overline{z}} \hspace{5mm}\textrm{ and
}\hspace{5mm} \hat{P}_{j} \equiv \sum_{z \in P_j }\ketbra{z}{z}
\end{eqnarray}
onto the eigenspaces of the two Hamiltonians $H_{0}$ and $H_{1}$,
\begin{eqnarray}
H_{0}\equiv\sum_{k=0}^{q(n)} F_k\hat{Q}_k \hspace{5mm}\textrm{ and
}\hspace{5mm} H_{1}\equiv\sum_{j=0}^{p(n)} E_j\hat{P}_j,
\end{eqnarray}
where the two sequences of energy eigenvalues
$(E_j)_{j=0,1,\ldots,p(n)}$ and $(F_k)_{k=0,1,\ldots,q(n)}$ are
strictly increasing sequences of real numbers. Assume further that
$F_0=0=E_0$ and that the starting state $\ket{u}$ is a ground state
of $H_{0}$, i.e. $0^n\in Q_0$.  We will also write
\begin{equation}
H_0 = \sum_{z\in\{0,1\}^n} F (z)
\ketbra{\overline{z}}{\overline{z}}\hspace{5mm}\textrm{ and
}\hspace{5mm} H_1 = \sum_{z\in\{0,1\}^n}E (z) \ketbra{{z}}{{z}}
\end{equation}
for eigenvalue functions $E (z)$ and $F (z)$ which are defined by
the partitions $\mathcal{P}$ and $\mathcal{Q}$ and the sequences
$(E_j)$ and $(F_k)$.

Let $s(t)$ be a smooth, nondecreasing, real function of time $t$
such that
\begin{equation}
s:[0,T]\longrightarrow [0,1],
\end{equation}
with $s(0)=0$ and $s(T)=1$, for some final time $T$.  It is clear
that
\begin{equation}
H_{\textrm{gen}}(t) \equiv \left(1-s(t)\right)H_{0} + s(t)H_{1},
\hspace{10mm}0\leq t\leq T,
\end{equation}
is a Hamiltonian having suitable form for adiabatic quantum
algorithms solving problems for which $P_0$ contains the solutions.

\subsection{A general bound}\label{sec__ArbDrvHam}

\noindent Let $\ket{\psi_{\textrm{gen}}(t)}$ be the state evolving
under $H_{\textrm{gen}}(t)$, that is, we define the initial value
problem
\begin{eqnarray}
 \frac{\partial}{\partial t}\ket{\psi_{\textrm{gen}}(t)}&=&-\i
H_{\textrm{gen}}(t)\ket{\psi_{\textrm{gen}}(t)},\hspace{5mm} 0\leq t\leq T\\
\ket{\psi_{\textrm{gen}}(0)}&=&\ket{u}.
\end{eqnarray}
Let $K$ be the index of the largest set $Q_k$:
\begin{equation}
|Q_K|\geq |Q_k|\hspace{5mm}\textrm{for all $k=0,1,\ldots,q(n)$}.
\end{equation}
Note
\begin{equation}
H_{\textrm{gen}}(t) = \left(1-s(t)\right)\left(F_K\id
-\sum_{k=0}^{q(n)}(F_K-F_k)\hat{Q}_k\right) + s(t)\sum_z
E(z)\ketbra{z}{z}.
\end{equation}
Noting that
\begin{equation}
\ket{\overline{z}}\equiv W\ket{z} =\frac{1}{\sqrt{N}} \sum_x\sum_y
(-1)^{x\bullet y}\ketbra{x}{y}\ket{z}=\frac{1}{\sqrt{N}} \sum_x
(-1)^{x\bullet z}\ket{x},
\end{equation}
we have, dropping the time-dependence notation,
\begin{eqnarray}
\nonumber H_{\textrm{gen}}
&=&(1-s)\left(F_K\id-\frac{1}{N}\sum_x\sum_y\left(\sum_{k=0}^{q(n)}(F_K-F_k)\sum_{z\in
Q_k}(-1)^{z\bullet \left(x\oplus
y\right)}\right)\ketbra{x}{y}\right)\\
\nonumber &&\hspace{10mm}+ \hspace{2mm}s\sum_{z}E(z)\ketbra{z}{z}.
\end{eqnarray}

Suppose $\ket{\psi_{{\textrm{gen}}}(t)}$ is expressed in the
computational basis as $\ket{\psi_{{\textrm{gen}}}(t)} =
\sum_{z=0}^{N-1}\gamma_{z}(t)\ket{z}$. Clearly, the goal of the
unitary evolution is to make $\sum_{w'\in
P_0}|\braket{w'}{\psi_{\textrm{gen}}(T)}|^{2}=\sum_{w'\in P_0
}|\gamma_{w'}(T)|^{2}$ of constant (or just $1/\textrm{poly}(n)$)
order so that we have a good probability of discovering one of the
solutions $w'$ by performing a measurement in the computational
basis on the final state $\ket{\psi_{\textrm{gen}}(T)}$.  The
Schr\"{o}dinger equation implies
\begin{eqnarray}
\nonumber \frac{\partial}{\partial t} \ket{\psi_{\textrm{gen}}} 
&=&-i \sum_{x} \left( \sum_{y}\bra{x}H_{{\textrm{gen}}}\ket{y}
\gamma_y \right)\ket{x}.
\end{eqnarray}
For each solution $w'\in P_0$,
\begin{eqnarray}
\nonumber && \frac{\partial}{\partial t}|\gamma_{w'}|^{2}\\
\nonumber &=& 2\Re(  \bra{w'}\frac{\partial}{\partial t}
\ket{\psi_{\textrm{gen}}}\cdot\braket{w'}{\psi_{\textrm{gen}}}^{*} )
\\ \nonumber &=&
2\sum_{y}\bra{w'}H_{{\textrm{gen}}}\ket{y}\Im\left(\gamma_{w'}\gamma_{y}^{*}\right)\\
\nonumber &=& -2(1-s)\frac{1}{N}\sum_{y\neq
w'}\sum_{k=0}^{q(n)}(F_K-F_k)\left(\sum_{z\in Q_k}(-1)^{z\bullet
(w'\oplus y)
}\right)\Im\left(\gamma_{w'}\gamma_{y}^{*}\right)\\
\nonumber &=&
-2(1-s)\frac{1}{N}\sum_{y}\sum_{k=0}^{q(n)}(F_K-F_k)\left(\sum_{z\in
Q_k}(-1)^{z\bullet (w'\oplus y)
}\right)\Im\left(\gamma_{w'}\gamma_{y}^{*}\right),
\end{eqnarray}
where the last two lines hold because
$\Im(\gamma_{w'}\gamma_{w'}^*)=0$.  If the strings $z\in Q_k$ were
somehow ``random enough'', we would expect the sums
$\left(\sum_{z\in Q_k}(-1)^{z\bullet (w'\oplus y) }\right)$ to be
close to $0$. This would suggest that the derivative is small for an
average problem instance.  Without going into statistics about
typical problem instances, we can only bound the size of these sums
by $|Q_k|$. Using this bound for every $w'\in P_0$ and using
$\left|\sum_y \Im(\gamma_{w'}\gamma_y^*)\right|\leq\sqrt{N}$ (which
follows from applying the Cauchy-Schwarz inequality to the absolute
value of the dot-product of the vector of all 1's and the vector of
$\Im(\gamma_{w'}\gamma_y^*)$ values) we get
\begin{equation}
\left|\sum_{w'\in P_0}\frac{\partial}{\partial t}\abs{\gamma_{w'}}^2
\right|\leq 2\frac{|P_0|}{\sqrt{N}}\cdot\max_kF_k\cdot
\textrm{dim}\left(\textrm{span}\{\ket{\overline{z}}:z\in
Q_K\}^\perp\right).
\end{equation}
Integrating from $t=0$ to $t=T$ as before, in order to lower-bound
by a constant $c$ the total probability of finding a solution, we
require
\begin{equation}
T\geq \frac{c\sqrt{N} - |P_0|/\sqrt{N}}{2|P_0|\cdot\max_kF_k\cdot
\textrm{dim}\left(\textrm{span}\{\ket{\overline{z}}:z\in
Q_K\}^\perp\right)}.
\end{equation}
Assume that the number of solutions $|P_0|$ is small.  Thus if the
dimension of the subspace orthogonal to the largest eigenspace of
$H_0$ is small, then the algorithm requires exponential time $T$.

\subsection{Searching with a more complex initial
Hamiltonian}\label{sec__SrchWMoreCmplxDrvHam}

\noindent We can use a similar argument to prove the analogous
result for $H_1$.  In this section, we illustrate the main
ingredients of this similar argument by giving it in the simpler
case where the final Hamiltonian is a one-dimensional projector.
Thus, as explained below, we get an interesting variant of Farhi and
Gutmann's optimality result in Ref.~\refcite{FG96}.  We note that an
argument similar to that in Ref.~\refcite{FGGN05}, based on the
time-reversibility of quantum mechanics, can be used to prove the
following result; that is, the following result is implied by the
``time-reversal'' of the result in the previous section (or the
results in Ref.~\refcite{FGGN05} or Ref.~\refcite{WY06}, both of
which consider one-dimensional-projector initial Hamiltonians). Our
short proof below need not appeal to time-reversibility.

Consider the following problem: ``What is the minimum time $T$
needed such that the Hamiltonian
\begin{eqnarray}
H_{B}(t)&\equiv& H_{\textrm{gen}}\left(t\right)\left.\right|_{p(n):=1, |P_0|:=1}\\
&=&\left( 1-s(t)
\right)\sum_{z}F(z)\ketbra{\overline{z}}{\overline{z}} +
s(t)E_1\left( \id -\ketbra{w}{w}\right),\hspace{5mm} 0\leq t \leq T
\end{eqnarray}
(adiabatically) evolves the start state
$\ket{\psi\left(0\right)}=\ket{u}$ to a final state
$\ket{\psi\left(T\right)}$ that is close to $\ket{w}$?''  Noting
that
\begin{equation}
\ketbra{w}{w}=\frac{1}{N}\sum_x\sum_y (-1)^{w\bullet \left(x\oplus y
\right)}\ketbra{\overline{x}}{\overline{y}},
\end{equation}
we have
\begin{equation}
H_{B}=(1-s)\sum_{z}F(z)\ketbra{\overline{z}}{\overline{z}} +
sE_{1}\left(\id-\frac{1}{N}\sum_x\sum_y(-1)^{w\bullet \left(x\oplus
y \right)}\ketbra{\overline{x}}{\overline{y}}\right).
\end{equation}
Letting $\ket{\psi(t)} = \sum_z \beta_{z}(t)\ket{\overline{z}}$, the
Schr\"{o}dinger equation implies that
\begin{eqnarray}
\nonumber \frac{\partial}{\partial t} \ket{\psi} &=& -i
H_{B}\ket{\psi} =
-\i\sum_x\sum_y\bra{\overline{x}}H_B\ket{\overline{y}}\beta_y\ket{\overline{x}}.
\end{eqnarray}
Instead of bounding the rate at which probability amplitude flows
into the solution state $\ket{w}$, we bound the rate at which it
flows out of the start state $\ket{u}$.  Noting that
$\braket{u}{\psi(t)}=\beta_0(t)$, we can derive that
\begin{eqnarray}
\nonumber \frac{\partial}{\partial t}|\beta_0(t)|^{2} &=&
-\frac{2sE_1}{N} \sum_{y}(-1)^{w\bullet y }\Im(\beta_y\beta_0^{*}).
\end{eqnarray}
Using the Cauchy-Schwarz inequality we get the bound
\begin{equation}
\left|\frac{\partial}{\partial t}|\beta_0(t)|^{2}\right|\leq
\frac{2E_1}{\sqrt{N}},\hspace{5mm}0\leq t\leq T.
\end{equation}
Noting that $\beta_{0}(0)=1$ and using
\begin{equation}
\abs{\beta_{0}(T)}^2-\abs{\beta_{0}(0)}^2=\int_{0}^{T}\frac{\partial
\abs{\beta_{0}(t)}^2 }{\partial t} dt \geq
-\int_{0}^{T}\left|\frac{\partial \abs{\beta_{0}(t)}^2 }{\partial
t}\right| dt
\end{equation}
gives
\begin{equation}
|\beta_{0}(T)|^{2}\geq 1- \frac{2 E_{1} T}{\sqrt{N}}.
\end{equation}
The following inequality holds:\cite{Ioa02}
\begin{equation}\label{ineq__Zalka}
|\braket{w}{\psi_{B}(T)}|^2 + |\braket{u}{\psi_B(T)}|^2\leq
1+|\braket{u}{w}|.
\end{equation}
This implies
\begin{equation}
|\braket{w}{\psi_{B}(T)}|^2 \leq
\frac{2E_1T}{\sqrt{N}}+\frac{1}{\sqrt{N}},
\end{equation}
which implies that the algorithm requires exponentially large time
$T$ in order to compute $w$ with sufficiently high probability.

The result effectively bounds the power of any initial Hamiltonian
that is diagonal in the Hadamard basis, in the presence of a final
Hamiltonian that merely encodes the binary function $f(z)$ defining
the generalized search problem (with unique solution $w$). We
interpret this result in the context of Farhi and Gutmann's
continuous-time search algorithm.\cite{FG96} Recall that their
optimality result is that no Hamiltonian of the form $H_D(t) + H_w$
can solve the continuous-time search problem faster than $H_u + H_w$
if it must work faster on \emph{most} problem instances $w\in
\{0,1\}^n$. It follows from our result that no $H_D(t)$ \emph{that
is restricted to being diagonal in the Hadamard basis} can
outperform $H_u$ on \emph{even one} problem instance.

\section*{Conclusion}

\noindent Combining the techniques of Sections \ref{sec__ArbDrvHam}
and \ref{sec__SrchWMoreCmplxDrvHam}, we can show that, in our
generalized adiabatic scheme, both $H_0$ and $H_1$ must have at
least two large mutually-orthogonal eigenspaces if $T$ is to be
subexponential in $n$.

\section*{Acknowledgements} \noindent LMI acknowledges the
support of NSERC, the University of Waterloo, and EPSRC (UK).  MM is
supported by DTO-ARO, NSERC, CFI, ORDCF, CIAR, CRC, ORF, and
Ontario-MTI.

\end{document}